\documentclass[a4paper,linenos]{jpconf}
\usepackage{graphicx}
\usepackage{lineno}
\usepackage{hyperref}

\begin{document}
\title{Use of checkpoint-restart for complex HEP software on traditional
architectures and Intel MIC}

\author{Kapil Arya$^1$, Gene Cooperman$^1$, Andrea Dotti$^2$, Peter Elmer$^3$}

\address{$^1$ College of Computer and Information Science, Northeastern University, Boston, MA, USA}
\address{$^2$ SLAC National Accelerator Laboratory, Menlo Park, CA 94025, USA }
\address{$^3$ Department of Physics, Princeton University, Princeton, NJ 08540, USA}

\ead{Peter.Elmer@cern.ch}

\begin{abstract}
Process checkpoint-restart is a technology with great potential
for use in HEP workflows. Use cases include debugging, reducing the startup
time of applications both in offline batch jobs and the High Level Trigger,
permitting job preemption in environments where spare CPU cycles are being 
used opportunistically and efficient scheduling of a mix of multicore and 
single-threaded jobs. We report on tests of checkpoint-restart technology 
using CMS software, Geant4-MT (multi-threaded Geant4), and the 
DMTCP (Distributed Multithreaded Checkpointing) package. We analyze both 
single- and multi-threaded applications and test on both standard Intel 
x86 architectures and on Intel MIC. The tests with multi-threaded applications 
on Intel MIC are used to consider scalability and performance.  These are 
considered an indicator of what the future may hold for many-core computing.

\end{abstract}

\section{Introduction}

The computing requirements for high energy physics (HEP) projects like the 
Large Hadron Collider (LHC)~\cite{LHCMACHINE} at the European
Laboratory for Particle Physics (CERN) in Geneva, Switzerland are
larger than can be met with resources deployed in a single computing
center. This has led to the construction of a global distributing computing 
system known as the Worldwide LHC Computing Grid (WLCG)~\cite{WLCG},
which brings together resources from nearly 160 computer centers in 35 
countries. Computing at this scale has been used, for example, 
by the CMS~\cite{CMSDET} and Atlas~\cite{ATLASDET} experiments for the 
discovery of the Higgs Boson~\cite{CMSHIGGS,ATLASHIGGS}. To achieve this
and other results the CMS experiment, for example, typically used during 
2012 a processing capacity between 80,000 and 100,000 x86-64 cores from
the WLCG.
Further discoveries are possible in the next decade as the LHC moves to 
its design energy and increases the machine luminosity. However, increases in
dataset sizes by 2-3 orders of magnitude (and commensurate processing
capacity) will eventually be required to realize the full potential of this
scientific instrument.

Building the software to run on and operate such a computing system is 
a major challenge. The distributed nature of the system implies that 
ownership and control of the resources is also distributed, and thus
the resources are by necessity heterogeneous in nature. This heterogeneity
appears both in
terms of specific x86 hardware generations and in patch levels of the
deployed Linux operating systems. As these resources are at times shared
with other projects custom modifications of systems for HEP-specific or 
experiment-specific reasons are in general not possible.
The very large number of CPU hours used also introduces significant
reliability requirements on the software. The software itself is 
non-trivial: each experiment typically is dependent on many millions of 
lines of code, written for the most part in C++ with contributions from up
to a thousand physicists. Given the upgrade and evolution plans for the 
LHC, these software projects, begun in the late 1990's, will likely need 
to evolve with computing technology through the 2020's.

Such an environment however provides significant opportunities for
innovation. In this paper we examine one interesting technology, {\it process 
checkpoint-restart}, which has great potential for use in HEP workflows in 
such a system. We first describe the specific use cases of interest
and the requirements to make the technology useful in the HEP environment.
We then provide some benchmarks for the use of checkpoint-restart
with the CMS software on today's x86-64 processors. And finally, we examine 
aspects of this technology which are of interest given the possible evolution 
of processor technologies and resource availability in the coming years. 
We look in particular at the use of checkpointing with architectures
like Intel's Xeon Phi, a member of Intel's MIC (many integrated core)
architecture.

\section{Process Checkpoint-Restart}

  In many circumstances it is desirable to ``checkpoint'' the state of
a UNIX process, or a set of processes, to disk with the possibility of
restarting it at a later time. 

\subsection{Use Cases}
\label{sec:usecase}

There are a number of interesting use cases for this functionality:

{\bf Debugging:} The very large number of jobs and CPU hours required
for HEP computing makes high reliability of the software quite important. 
The distributed and heterogeneous nature of the computing system however 
makes debugging problems somewhat difficult, as the first step to resolving 
most code behavior problems is being able to reproduce the problem. While 
a traditional ``core'' file may provide information about the process
state after a crash has happened, it doesn't allow one to step through the
program to see the behavior leading to the crash. In the case where the
crash happens after a job has run for many hours, reproducing a problem
by rerunning it from the beginning can also be quite expensive.
  If however a job were to checkpoint its state from time to time, it
would be possible to use the last checkpointed state before a crash to 
reproduce and replay the problem quickly.

{\bf Avoiding CPU-intensive initialization steps:} Many applications
are constructed such that they have CPU-intensive initialization steps.
Examples include in-memory geometry construction from a simplified geometry
description, physics cross section table calculations, etc. 
Typically this is done to allow generality of software implementation
for multiple possible job configurations: it is easier to calculate
quantities derived from job configurations on the fly at the beginning
of each job than to store those quantities for a possibly infinite
number of potential job configurations. HEP workflows are however 
constructed such that a particular job configuration may be used for a 
very large number of jobs, where the geometry or physics process 
configuration are the same and only random number seeds or input files
change from job to job. The result is that the same calculation is
done in every single job instance in a particular workflow.
In most cases where jobs run for a long time this job initialization
time is negligible relative to the total running time.

However two cases exist where job initialization time can be problematic.
First, very short duration
jobs can sometimes be required for other software reasons or for 
operational reasons related to resource availability or input dataset
structure. In this case the overhead from long startup initialization
times can be a significant fraction of the CPU utilization. 

Second,
as will be described later, there are strong reasons to consider the
use of multi-threaded applications in the future. In the case where the
startup initialization itself cannot be easily parallelized and
will be executed sequentially on a single core, the initialization itself 
may effectively idle a large number of other cores eventually needed
for the event processing.
In this case, a single instance of the job can be run and checkpointed
just after the initialization phase. That checkpoint can then be used
to restart a much larger number of instances of the application in batch,
with only minor reconfiguration to set input files and/or random number
seeds. The instances running in batch can thus avoid the startup CPU cost.

{\bf Allowing preemption during opportunistic resource use:} From time
to time, it is possible to ``opportunistically'' use computing resources
which belong to some other
organization  when the other organization does not 
have enough work to keep the processors fully utilized for some period of 
time. In some of these cases the period of time for which the resources
are made available may not be well defined in advance and the resources may
need to be handed back to their owner in an unscheduled fashion. In this
case it is useful to be able to ``preempt'' running opportunistic jobs,
checkpoint their state to disk and restart them when opportunistic use
is again possible.

{\bf Interactive ``workspaces'':} In interactive programs, such as
event displays and analysis tools, the user provides inputs which 
lead to particular state of the program at a given time. Being able
to save that state out, for example before going home for the day,
and restart later is often desirable.

{\bf Very long running jobs:} In situations where jobs must run for
an extremely long time, sometimes days or weeks, they can be sensitive to 
hardware or infrastructure failures or interfere with required site
maintenance. In these cases it can be quite useful to checkpoint periodically 
the program to avoid losing and needing to repeat the calculations from
scratch after such failures.

{\bf Managing ``tails'' for multi-threaded applications}: Several HEP
experiments are moving in the direction of multi-threaded frameworks,
which (initially) process events on different threads. As the CPU
time per event can vary significantly (with long ``tails'' to
the distribution) at the end of the job, one thread may still be processing
an event which takes a long time while the other threads/cores are idled.
One possibility for managing such situations would be to checkpoint the
job with a single active thread and restart a number of such jobs at
a later time together, to keep the full set of CPU cores active.

\subsection{Possible solutions}
  A rudimentary ``checkpoint-restart'' can sometimes be achieved in an 
application-specific fashion. For example a typical HEP event processing 
framework can be constructed to perform a simple ``checkpoint'' by
flushing completed output events to its 
output file after every N input events have been processed. In addition 
it is necessary to write
some sort of ``metadata'' to track any other relevant internal state
needed to restart the job, e.g. how far the job had progressed through
its input events, the state of random number generators, etc.
In this example a ``restart'' would then be performed by restarting the
framework and passing it information to allow it to reconfigure itself
to match the state it was in at the time of the output checkpoint.
This requires however the addition and maintenance of dedicated code, both 
in the framework itself and externally in the workflow management system. In
some cases, where third party libraries are used which also maintain
state, it can be quite complex to truly restore the same state.
If the state is encapsulated within the code of the library, for software
engineering reasons, it can also be impossible.

  A much better, and more general, solution is true process-level
checkpointing. This is a technology which has existed for a long time,
especially in High Performance Computing (HPC) and batch systems,
however often the particular implementations are tied
to specific environments. Thus the technology has not seen general
use in HEP high throughput distributed computing. In this paper we examine 
the use of a transparent, user-level checkpointing package for distributed 
applications called Distributed MultiThreaded 
CheckPointing (DMTCP)~\cite{DMTCP}. The features of DMTCP make it more 
appropriate for deployment in the HEP distributed computing environment.

\section{Distributed MultiThreaded CheckPointing (DMTCP)}

DMTCP (Distributed MultiThreaded CheckPointing) is free, open source
software (\url{http://dmtcp.sourceforge.net}, LGPL license).  The DMTCP
project traces its roots to late~2004. A key feature of DMTCP for use
in the heterogeneous HEP computing environment is
that it works in user space, with no kernel-level modifications required.
As such it is works with a wide range of Linux kernel versions. It also works
with multi-threaded applications and compression of output checkpoint files
is possible.

Its usage can be as simple as:
{\tt
\begin{verbatim}
  dmtcp_launch ./myapp arg1 ...
  dmtcp_command --checkpoint    [ from a second terminal window ]
  dmtcp_restart ckpt_myapp_*.dmtcp
\end{verbatim}
}

DMTCP is also ``contagious''.  If a process begins under DMTCP control, then
any child processes will also be under DMTCP control, and any remote
processes (spawned through ``ssh'') will also be under DMTCP control.
At the time of checkpoint, a script, {\tt dmtcp\_restart\_script.sh},
is written, and the script can restart all processes across all
nodes for the given computation.

The newly released DMTCP version~2.0 (as of Oct. 3, 2013), supports DMTCP
plugins to flexibly adapt to external conditions.  For example,
the DMTCP plugin interface permits application-initiated
checkpoints, as well as application-delayed checkpoints during critical
operations.  Alternatively, the {\tt --interval} flag of
{\tt dmtp\_launch} permits automatic periodic checkpointing.

DMTCP plugins make it easier to use event hooks to detach such external
resources as a database prior to checkpoint, and to reconnect during
restart.  While DMTCP will save and restore the file offset of open
files, event hooks make available an alternative of cleanly closing
valuable files during checkpoint, and re-opening them during restart.
In another example, DMTCP virtualizes network addresses to enable
transparent migration to a new cluster.  Finally, if a large region of
memory is not actively used at the time of checkpoint, then the size of
the checkpoint image can be considerably reduced.  An event hook allows
the application to write zeroes into the inactive memory at checkpoint
time, and DMTCP will then replace the zeroes by zero-fill-on-demand pages
(empty pages to be recreated on demand).

Among the contributed plugins for DMTCP is the rm (resource manager)
plugin, which supports use of DMTCP within the Torque and SLURM
batch queues.  Plugin support for additional batch queues is planned.
Similarly, future support for the Intel SCIF network is planned,
allowing one to checkpoint a computation over a network of Intel MIC
CPUs.  The SCIF plugin will be based on the existing contributed plugin
to support checkpointing over InfiniBand.  Other contributed plugins
support checkpointing a network of virtual machines.  Virtual machines
ease the job of deploying complex software.

\section{Experimental results with x86-64}

To investigate the characteristics of DMTCP with HEP software, we chose
to make tests using the CMS simulation application. The machine
used to do the test was a dual quad-core Intel Xeon L5520 operating 
at 2.27GHz, with 24GB of memory. Tests were done with checkpoint files
written to a local disk as well as the normal job output files. The
test job had no input event file. However it reads conditions via
a web squid from a remote database. DMTCP version 1.2.7 and CMSSW version 
$6\_2\_0$, compiled with GCC version 4.7.2, were used for the tests. For
simplicity the test job used
with checkpointing was the only CPU-intensive process running on the 
machine at the time the tests were performed. The tests were done using 
the external DMTCP coordinator trigger to produce a checkpoint, rather 
than the API.

The CMS application generated simple Minimum Bias events and
simulated them with Geant4~\cite{GEANT4}. The particular test job itself 
has a 2 minute initialization phase and then takes an average time per 
event of $\sim$13 seconds. The memory footprint is $\sim$1GB VSIZE (750MB RSS).
A typical uncompressed checkpoint takes $\sim$1.5s and the resulting
size on disk of the checkpoint file was $\sim$750MB. When triggering 
checkpointing with the compression on, $\sim$10s was required. 
The checkpoint image was however significantly smaller, at only
$\sim$220MB. In both cases no 
problems were seen in restarting the application from the checkpoint files.

\section{Processor Architectures}

  The construction of the WLCG was greatly facilitated by the convergence
around the year 2000 on commodity x86 hardware and the standardized
use of Linux as the operating system for scientific computing clusters. 
Even if multiple generations of x86 hardware (and hardware from both Intel 
and AMD) are provided in the various computer centers, this was a far
simpler situation than the previous typical mix of proprietary UNIX operating
systems and processors. Until around 2005, a combination of increased 
instruction level parallelism and (in particular) processor clock frequency 
increases insured that performance gains expected from Moore's Law would 
be seen by single sequential applications running on a single processor.
The combination of Linux, commodity x86 processors and Moore's Law gains
for sequential applications made for a simple software environment.

However since around 2005 processors have hit scaling limits, largely driven 
by overall power consumption~\cite{GAMEOVER}. The first large change in 
commercial processor products as a result of these limits was the 
introduction of ``multicore'' CPUs,
with more than one functional processor on a chip. At the same time
clock frequencies ceased to increase with each processor generation and 
indeed were often reduced relative to the peak. The result of this was
one could no longer expect that single, sequential applications would
run faster on newer processors. However in the first approximation,
the individual cores in the multicore CPUs appeared more or less
like the single standalone processors used previously. Most large
scientific applications (HPC/parallel or high throughput) run in
any case on clusters and the additional cores are often simply
scheduled as if they were additional nodes in the cluster. This
allows overall throughput to continue to scale even if that of a
single application does not. It has several disadvantages, though,
in that a number of things that would have been roughly constant
over subsequent purchasing generations in a given cluster (with
a more or less fixed number of rack slots, say) now grow with each
generation of machines in the computer center. This includes the
total memory required in each box, the number of open files and/or
database connections, increasing number of independent (and incoherent)
I/O streams, the number of jobs handled by batch schedulers,
etc.  The specifics vary from application to application, but
potential difficulties in continually scaling these system parameters
puts some pressure on applications to make code changes in response,
for example by introducing thread-level parallelism where it did
not previously exist.

There is moreover a more general expectation that the limit of power
consumption on future Moore's Law scaling will lead to more profound
changes going forward. In particular, the power hungry x86-64
``large'' cores of today will likely be replaced by simpler and less power 
hungry ``small'' cores. One example of such a technology is the Intel
MIC architecture, as implemented in the Intel Xeon Phi coprocessor card.

\section{Experimental Results with Xeon Phi}
  
To test the use of checkpointing on the Xeon Phi, we used a beta version of Geant4 version 10 which provides support for event-based multi-threaded applications. We did not use the full CMS simulation for this, but instead a simpler benchmark application (FullCMS) which uses the actual CMS geometry imported from GDML file. The experimental setup was a standard Intel Xeon box with 32 logical cores, equipped with an Intel Xeon Phi 5120P coprocessor card.
In our test the application is started and checkpoints are triggered in different moments. Then we have restarted the application from the checkpoint file and we have verified that the application resumes correctly from the saved state. This condition is verified checking that the final output of the simulation equals the original one. A comparison of random number engine status at the end of the job. Since in a Geant4 application there is a very large use of random numbers (billions of calls to the engine in typical application), if the status of the application from the checkpoint file does not match exactly the original one the sequence of the random number calls will be different, producing a different final state of the random number engine. 

\begin{figure}[tbp]
\centering
\includegraphics[width=0.8\textwidth]{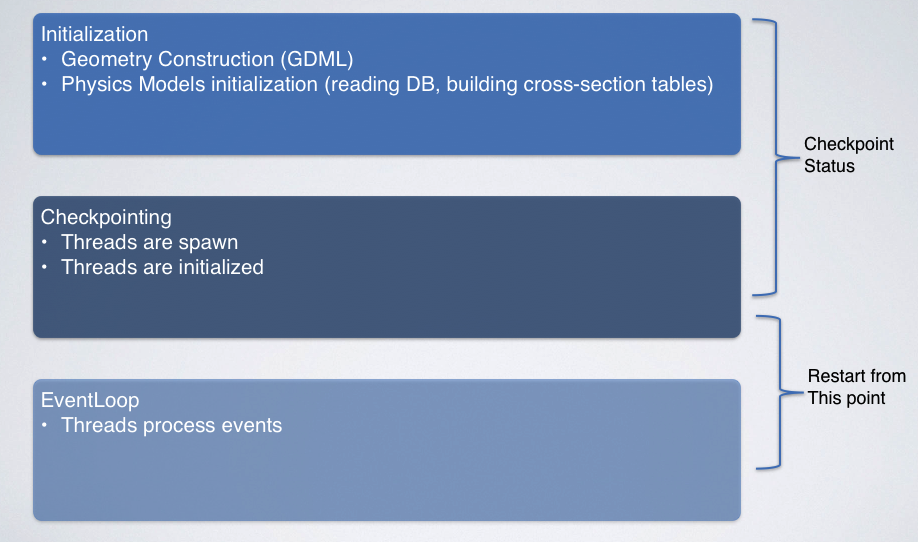}
\caption{Startup structure of a typical Geant4 simulation job}
\label{fig:geant4-ckpt}
\end{figure}

A typical Geant4 application with multi-thread support consists of a sequential part in which the geometry of the experimental setup is built in memory and physics processes are initialized (e.g. material-dependent cross sections are calculated). Threads are then spawned, initialized and they start to simulate events independently (see Figure~\ref{fig:geant4-ckpt}). To reduce the total memory footprint the most memory consuming objects are shared between threads. The need for synchronization between threads (locks, barriers) is minimized since only read-only objects are shared.

We have performed tests to verify the correct behavior of DMTCP for two of the Use Cases described in Section~\ref{sec:usecase}.

\begin{figure}[tbp]
\centering
\includegraphics[width=0.8\textwidth]{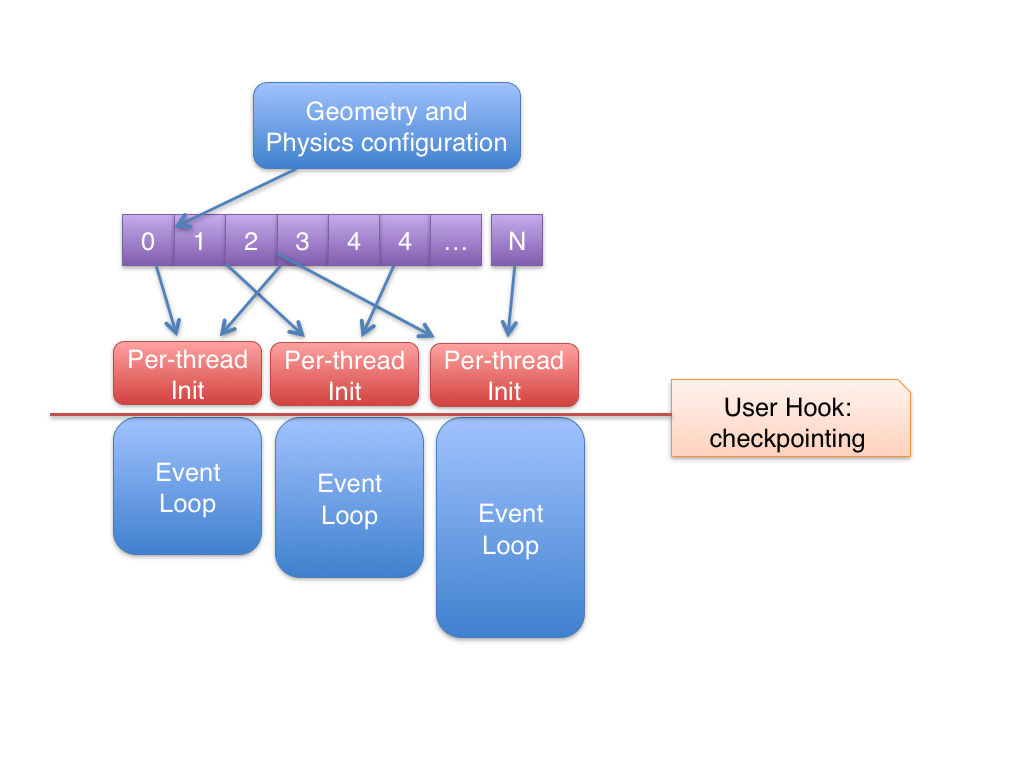}
\caption{User hook for checkpointing to avoid startup CPU cost}
\label{fig:ckpt1}
\end{figure}

In the first case we have instrumented the Geant4 application code with a call to DMTCP to trigger a checkpoint file at the end of the initialization phase (Figure~\ref{fig:ckpt1}). On the Intel Xeon Phi accelerator the initialization takes about 5 minutes. The checkpointing itself takes about 1 minute (the application working directory, physically located on the host, was mounted by the coprocessor through NFS) and the resulting checkpoint image file is 1.4GB (uncompressed). Restarting from the checkpoint image file takes less than 10s. Functionally this appears to work as expected. The resulting checkpoint image file can thus be distributed to other nodes and the simulation process ``cloned'' without the startup cost, simply by resetting the random number seeds. 

\begin{figure}[tbp]
\centering
\includegraphics[width=0.74\textwidth]{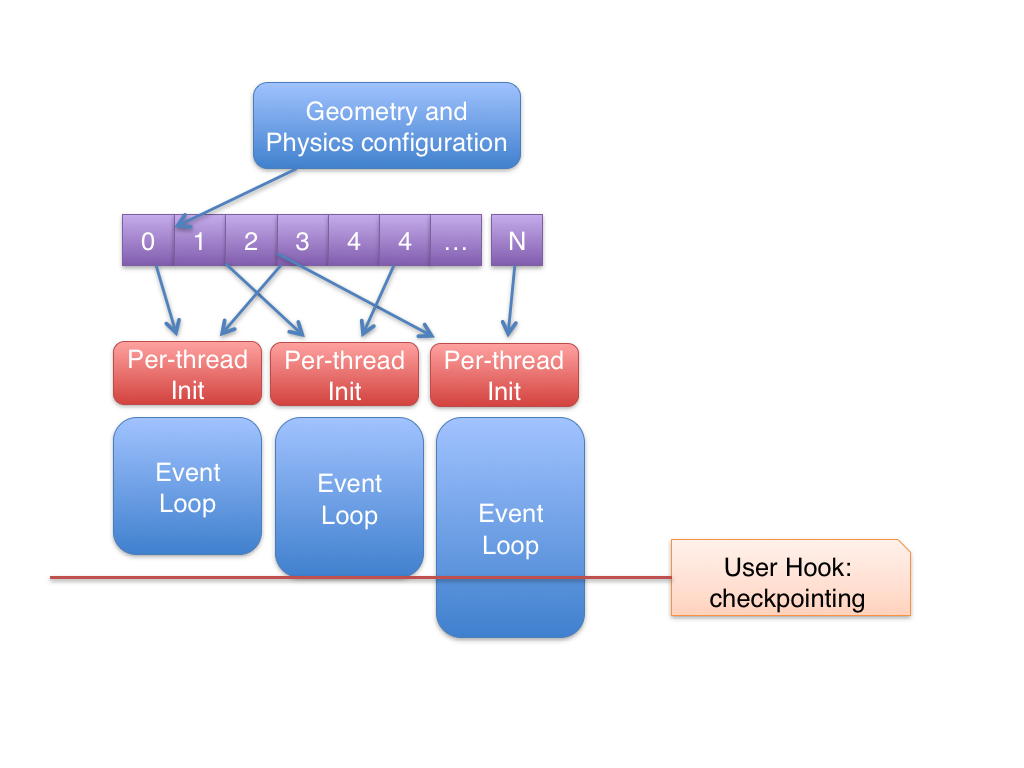}
\caption{User hook for checkpointing to avoid ``tails'' CPU cost}
\label{fig:ckpt2}
\end{figure}

For the second test we have emulated the use case of some threads being slower than other in producing results. This  can happen if one or more threads is simulating more complex-than-average events. To control such behavior we have modified our application in a way that half of the threads were responsible of simulating simple and fast events (low energy single particles) while the second half of threads was responsible for longer ones (ten times higher energy). In current Geant4 multi-threaded mode the application will wait for all threads before terminating thus leaving half of the MIC cores unused. We have instrumented the application code to verify when the number of active threads drops below a given value, in such a case a checkpoint is triggered (Figure~\ref{fig:ckpt2}). Also in this case we have verified that the application was behaving as expected.

It is important to note that we did not have to modify Geant4 ``kernel'' code to enable checkpointing, all code modifications were done at the application level. For the first test (checkpointing at first event) we have provided feedback to Geant4 developers that have introduced a new user-hook, not present in the initial design of the code, that allows for the execution of (optional) user code just before the first event is simulated but after all threads have been fully initialized (this guarantees that the checkpointing is performed in a reproducible state of the application). 
Both tests show that checkpointing can be used to increase the efficiency of resource usage also on accelerator technologies where the minimization of the time spent in sequential fractions of the code or with only few threads active is fundamental to efficiently use the hardware resources.

\section{Conclusions}
We have made investigations into the use of the checkpoint-restart
technology DMTCP with HEP applications from CMS and Geant4. We
have reported on the performance seen, both on a traditional x86-64
architecture and on Intel's Xeon Phi, for situations relevant for
a number of interesting use cases for HEP computing. We believe
that the results obtained are very encouraging and demonstrate the
viability of the use of this technology in the HEP environment.

\section*{Acknowledgments}
This work was partially supported by the National
Science Foundation under Grant~OCI-0960978 (Arya, Cooperman) and
Cooperative Agreement PHY-1120138 (Elmer)
as well as by the U.S. Department of Energy (Dotti).


\end{document}